\documentclass[12pt,a4paper]{article}
\usepackage[utf8]{inputenc}%agregado
\usepackage{amsmath}
\usepackage{graphicx}
\usepackage{amssymb}%agregado
\usepackage{amsthm}%agregado
\usepackage{color}	

\usepackage[breaklinks=true,backref=page]{hyperref}

\usepackage[english]{babel}%agregado
\usepackage[square,numbers]{natbib}%agregado ref apa

\usepackage[nottoc]{tocbibind}%poner referenica en la tabla de contenidos

\usepackage{multirow}%para unir filas

\usepackage{amsthm}
\theoremstyle{plain}% default

\theoremstyle{definition}

\theoremstyle{remark}

\title{Modelling macroparasitic diseases dynamics}
\author{Gonzalo Maximiliano LOPEZ$^{1,3,4}$, Juan Pablo APARICIO$^{1,2}$\\
	\\
	{\small $^1$ Instituto de Investigaciones en Energ\'ia no Convencional (INENCO),} \\ {\small Consejo Nacional de Investigaciones Cient\'ificas y T\'ecnicas (CONICET),}\\
	{\small Universidad Nacional de Salta, Av. Bolivia 5150, 4400 Salta, Argentina.}\\
	$^2${\small Simon A. Levin Mathematical, Computational and Modeling Sciences Center,} \\ {\small Arizona State University, PO Box 871904 Tempe, AZ 85287-1904, USA}\\
	{\small $^3$ Departamento de Matem\'atica, Facultad de Ciencias Exactas,}\\{\small Universidad Nacional de Salta, Av. Bolivia 5150, 4400 Salta, Argentina.}\\
	{\small $^4$ Corresponding author: gonzalo.maximiliano.lopez@gmail.com}}
\date{}

\usepackage{bm}

\usepackage{subfig}%para el subfloat

\usepackage{float}
\floatstyle{plaintop}
\restylefloat{table}

\newcommand{\nc}{\newcommand}
\nc{\R}{\mathbb{R}}
\nc{\N}{\mathbb{N}}
\nc{\pa}{\partial}
\nc{\F}{\mathfrak{H}}
\nc{\f}{\mathfrak{f}}

\begin{document}
\maketitle
\begin{abstract}
	\addcontentsline{toc}{section}{Abstract}
	
	In this work we present a general framework for the modeling of the transmission dynamics of macroparasites which do not reproduce within the host like \textit{Ascaris lumbricoides}, \textit{Trichuris trichiura}, \textit{Necator americanus} y \textit{Ancylostoma duodenale}.

	The basic models are derived from general probabilistic models for the parasite density-dependent mating probability. Here we considered the particular, and common case, of a negative binomial distribution for the number of parasites in hosts. We find the basic reproductive number  and we show that the system exhibit a saddle-node bifurcation at some value of the basic reproduction number. 
	 We also found the equilibria and basic reproduction number of a model for the more general case of heteregeneous host populations.

%	{\color{red}
%	
%	
%	In this paper, we study the transmitted dynamics of macroparasite. 
%	
%	We formulate and analyse  deterministic modeles using nonlinear differential equations. 
%	
%	The basic reproduction number is obtained  and endemic equilibrium points are shown to be asymptotically stable under given threshold conditions. 
%	}

\end{abstract}
\tableofcontents
\tableofcontents
\section{Introduction}
	
 	Mathematical models play an important role in understanding the transmission and impact of macroparasite diseases control measures \cite{anderson1992infectious,anderson2014coverage,truscott2016soil}.

	The first works on the theory of helminth infection was published in the 1960s by Tallis and Leyton by developing stochastic models of nematode parasite transmission in sheep and cattle \cite{leyton1968stochastic,tallis1966stochastic,tallis1969stochastic}.

 	Simultaneously Macdonald show that a consequence of sexual reproduction of distributed parasites within individual hosts was the inability to generate fertile infectious material when prevalence is low \cite{macdonald1965dynamics}.

 	Anderson and May then introduced much more general descriptions of helminth population dynamics. They developed descriptions for a model based on host age, distribution of parasite numbers per host, density dependence of egg production, and sexual mating functions that depend on parasite distribution and reproductive habits \cite{anderson1982population,anderson1992infectious}.
 	
 	%{\color{red}
 	
 	In this article we develop an analytical framework to describe the transmission dynamics of most macroparasite infections.
 	We first describe the dynamics of infection transmission by macroparasites. We then present two deterministic models for these transmission dynamics, the first for a homogeneous host community and the second for a heterogeneous host community.

 	In both models, reproductive characteristics of the parasite are considered, such as egg production and mating probability, both modeled by the density-dependent fecundity of the parasite and the distribution of parasites per host, which we assume to be negative binomial.

 	For both models we present the calculations of the equilibrium values and the basic reproduction number $R_0$ defined for the case of macroparasites as the average number of new parasite offspring caused by a typical parasite, from one generation to the next.
 	Finally for the homogeneous model we show that it has a saddle-node bifurcation.
 	
 	%}

\section{General framework}

Microparasite diseases are usually modeled using compartmental models. After infection,  microparasite population may rapidly grow into the host. This intra-host parasite dynamics determines the level of infectiousness of the individual. In a simple compartmental  model like the $SIR$-model  all the susceptible individuals are grouped in one class of size $S$, all the infected and infectious indviduals in a class of size $I$ and all the recovered individuals in a class of size $R$. Many refinements are possible, but the evolution of the parasite population within the host it is not considered or very simplified (for models including intra-host population dynamics 
%{\color{red}
see for example \cite{gandolfi2015epidemic})
%Gandolfi, A., Pugliese, A., \& Sinisgalli, C. (2015). Epidemic dynamics and host immune response: a nested approach. Journal of mathematical biology, 70(3), 399-435. and references therein.)}. 
The most common refinement consists in dividing infected individuals in two classes, exposed (those infected but not infectious yet) and infectious which leads to the well known $SEIR$ type models. 

\begin{figure}[t!]
	\centering
	\includegraphics[width=0.99\linewidth]{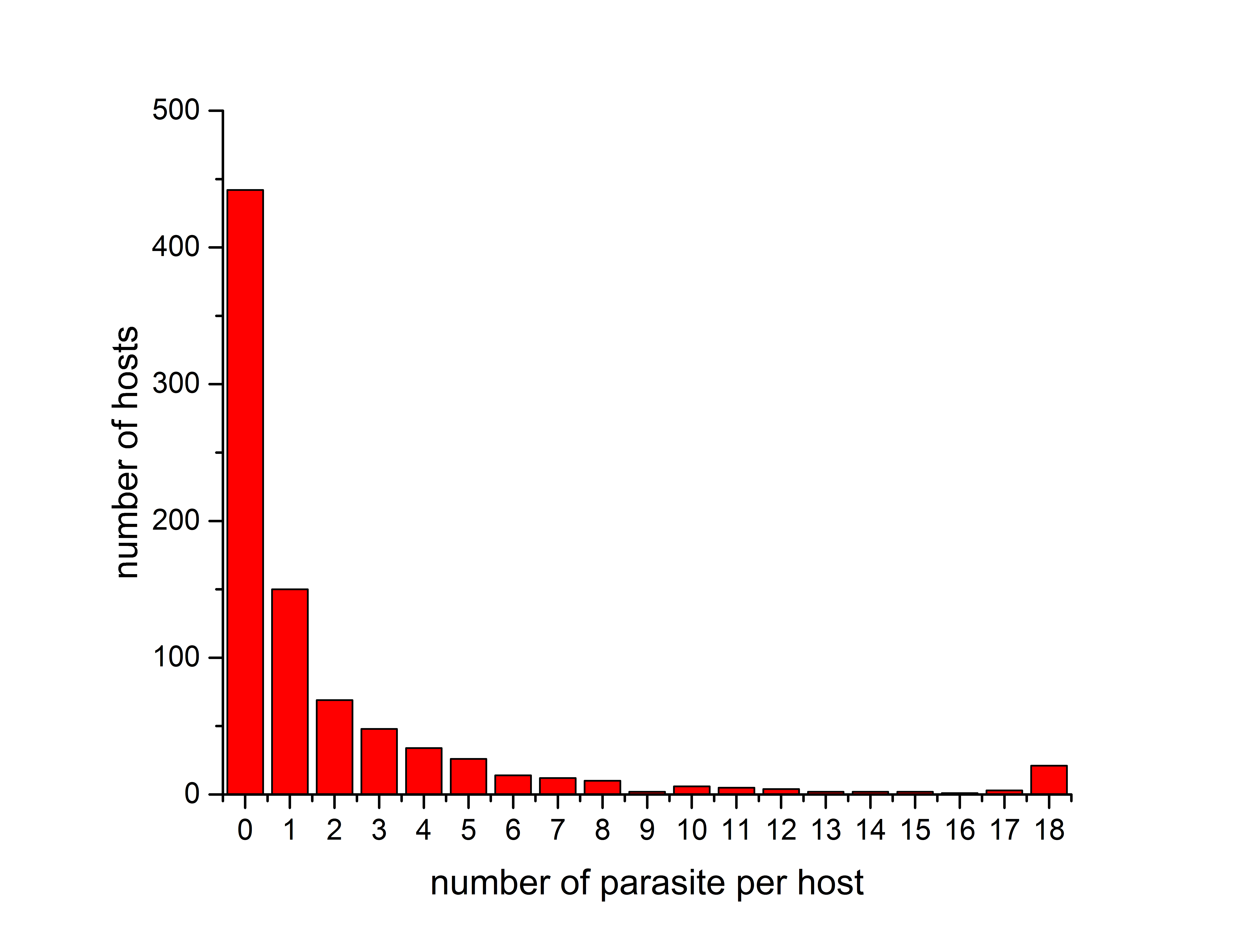}
	\caption{Distribution of \textit{Ascaris lumbricoides} parasite numbers per host in a study in rural populations in Korea \cite{seo1979frequency}. 
	Most hosts are uninfected or infected with a low burden of parasites while few are infected by large numbers of parasites.}
	\label{fig:dataseo}
\end{figure}

For most macroparasites the situation is completely different as these type of parasites do not reproduce within the host.  Most  infected  individuals have few macroparasites with a non-bell shaped distribution (see Figure \ref{fig:dataseo}) where few individuals concentrate most of the parasites in the host population \cite{seo1979frequency,lopez2022simple}. Negative binomial distributions usually provide a good description of the data. On the other hand, there is no host-to-host transmission of macroparasites as life cycle completes in the environment (from where host get infected).

Therefore the number of infected hosts it is not a representative variable of the parasite burden. Simple models for macroparasites consider the evolution of the mean burden of parasite within the population as well as the environmental parasite reservoir (which is composed by eggs or larvae). From the mean burden, the total parasite population is easily estimated.

\section{A basic model}
\label{s:basicmodel}
 
\subsection{Model structure}
\label{ss:structure}
The model presented in this paper is based on a model developed by Anderson and May \citep{anderson1992infectious,anderson1985helminth}.
The conceptual framework of parasite transmission dynamics is conceptualized as a population of mature parasites within human hosts and a population of infective stages (eggs or larvae) found in the environment (reservoir).
Hosts can become infected by contact with the infective stages (eggs or larvae) and can contaminate the environment (reservoir) with infective stages (eggs or larvae).

In a simple model for transmission dynamics of macroparasites in a population (where host demography is ignored) of size $N$ of hosts 
the dynamic variables  are the mean parasite burden of the population, $m$, and the infective stages in the environment formed by eggs or larvae, $\ell$.

In the following we will sketch the procedure to find parasite-related parameters from a statistical-probabilistic model for the parasite population.

The environmental parasite reservoir, composed by eggs or larvae, increases due to the contribution of adult parasites within the hosts. As most host harbor only few parasites, only hosts with at least one female and one male parasites will contribute with fertilized eggs to the reservoir. We will consider that the random variable $W$, the number of parasites in a host follow a negative binomial distribution. 
Therefore, the probability of observing $n$ parasites in a host is
\begin{equation}\label{disnb}
\mathrm{P}(W=n)=\frac{\Gamma(k+n)}{\Gamma(n+1)\Gamma(k)}\left( \frac{k}{m+k}\right) ^{k} \left( \frac{m}{m+k}\right) ^n
\end{equation}
where $m$ is the mean value (the mean population parasite burden) and $k$ the shape parameter. 
%The variance decreases with $k$ as $igma^2=m+m^2/k$. 
The variance increases with the reciprocal of $k$ as $\sigma^2=m+m^2/k$.

Mean egg production depends of the number of parasites within the host, it is a density-dependent process. 
A simple model for the average female fecundity of a female parasite in competition with $n-1$ parasites is given by 
$$\lambda(n)=\lambda_0 z^{n-1}$$ where $z=e^{-\gamma}$, and $\gamma$ is a parameter quantifying the intensity of the competition \cite{hall2000geographical}. 

Using the parasite host distribution \ref{disnb} we may compute the mean egg production per host as (\cite{lopez2022general})
$\lambda_0
%\dfrac{ m}{2} 
\alpha m
\psi(m,k,z)$
where $\alpha$  is the fraction of female parasites in a host and $\psi$ is given by
\begin{equation}
\psi(m,k,z)=\left[ 1+(1-z)\dfrac{m}{k}\right]^{-(k+1)}
\end{equation}
 is  known as the effective contribution of the female population to the parasite reservoir (in the form of eggs or larvae) \cite{churcher2006density}.

However only hosts with at least one female and one male parasites will effectively contribute to the parasite's reservoir by laying fertilized eggs. Therefore the mean fertilized egg production per host is 
\begin{equation}\label{meanfertilizedeggs}
\lambda_0
%\dfrac{ m}{2} 
\alpha m
\psi(m,k,z)\phi(m,k,z)
\end{equation} 
where $\phi(m,k,z)$ is the mating probability for the negative binomial distribution computed in (\cite{lopez2022general})
 \begin{equation}
 \phi(m,k,z)=1-\left[ \frac{1+ \left( 1-
 	%\dfrac{z}{2} 
 	\alpha z
 	\right) \dfrac{m}{k}}{1+(1-z)\dfrac{m}{k}}\right] ^{-(k+1)}
 \end{equation}

Therefore the mean fertilized egg contribution to the environmental reservoir  per host and per unit of time is
$\rho\lambda_0
%\frac{m}{2} 
\alpha m
\psi(m,k,z) \phi(m,k,z)$ where $\rho$ 
is the host's own contribution rate and the total contribution of eggs to the reservoir per unit of time of a host population $N$ is 
$\rho\lambda_0
%\frac{m}{2} 
\alpha m
\psi(m,k,z) \phi(m,k,z) N$. 
%{\color{red} tenemos que poner $\rho$? o ya esta eso en $\lambda(n)$?, verlo)} 
The population of eggs or larvae in the environment ($\ell$) also decreases due to egg/larval mortality ($\mu_{\ell}$) or due to host infection at the rate $\beta \ell$ per host. 

Therefore the dynamics of the reservoir is given by
	\begin{equation}\label{eqreservorio}
	\dfrac{d\ell}{dt}= \rho\lambda_0
	%\frac{m}{2} 
	\alpha m
	\psi(m,k,z) \phi(m,k,z) N- \mu_{\ell} \ell -\beta N \ell 
	\end{equation}

Finally, the dynamics for the mean burden $m$ is obtained as follow. Parasites are taken from the environment at the rate 	$\beta N \ell $ and therefore the mean burden increases at the rate  $\beta  N\ell/N=\beta\ell $. Parasites within the host die at the rate $\mu_p$ and hosts at the rate $\mu_h$ (killing all their parasites). Thus, the dynamics of $m$ is
\begin{equation}\label{eqm}
	\dfrac{dm}{dt}=\beta \ell - (\mu_h+\mu_p)m
	\end{equation}
	
	The system \eqref{eqreservorio}-\eqref{eqm} is the basic model of the transmission dynamics of macroparasites in a population of hosts.

\subsection{Equilibria and basic reproduction number}
	%Considerando el analisis de estabilidad para el sistema anterior y suponiendo una situación de equilibrio para el reservorio $L$, de la ecuación (\ref{model1eq2}) resulta
	%De la ecuación (\ref{model1eq2}) obtenemos que en el equilibrio
	From the equation \eqref{eqreservorio} we obtain that in equilibrium
	\begin{equation}\label{eqL}
		\ell^*=\frac{\rho N \lambda_0 \alpha}{(\mu_{\ell}+\beta N)} m \psi(m)\phi(m) 
	\end{equation} 
	%y reemplazado esto en la ecuación (\ref{model1eq1})	obtenemos la siguiente ecuación para la dinámica de $M$ 
	and substituting \eqref{eqL} in the equation \eqref{eqm} we get the following equation for the dynamics of $m$
	\begin{align}\label{eqMR0}
		\dfrac{dm}{dt}&=(\mu_h + \mu_p)\left[ R_0  \psi(m)\phi(m) -1 \right] m%\notag es para no tener numero en ecuacion
	\end{align}
	%donde el parámetro $R_0$ es el número reproductivo básico que es independiente de los efectos de la denso-dependencia   
	where the parameter $R_0$ is the basic reproductive number which, by definition, is independent of the effects of density-dependence and  mating probability
	\begin{equation}\label{valorR0}
	R_0=\frac{ N \lambda_0 \alpha  \rho \beta}{ (\mu_{\ell}+\beta N) (\mu_h + \mu_p) }
	\end{equation}
	where for a large $N$ value $R_0\approx \frac{ \lambda_0 \alpha  \rho }{ (\mu_h + \mu_p) }$.
	%Por lo tanto de la ecuación (\ref{eqMR0}) podemos obtener la condición de equilibrio para la carga media como 

	Therefore from the equation \eqref{eqMR0} we can obtain the equilibrium condition for the mean parasite burden
	\begin{equation}\label{eqequilibrio}
	\psi(m^*,k,z)\phi(m^*,k,z)=1/R_0
	\end{equation}
	
\begin{figure}[h!]
		\centering
		\includegraphics[width=0.99\linewidth]{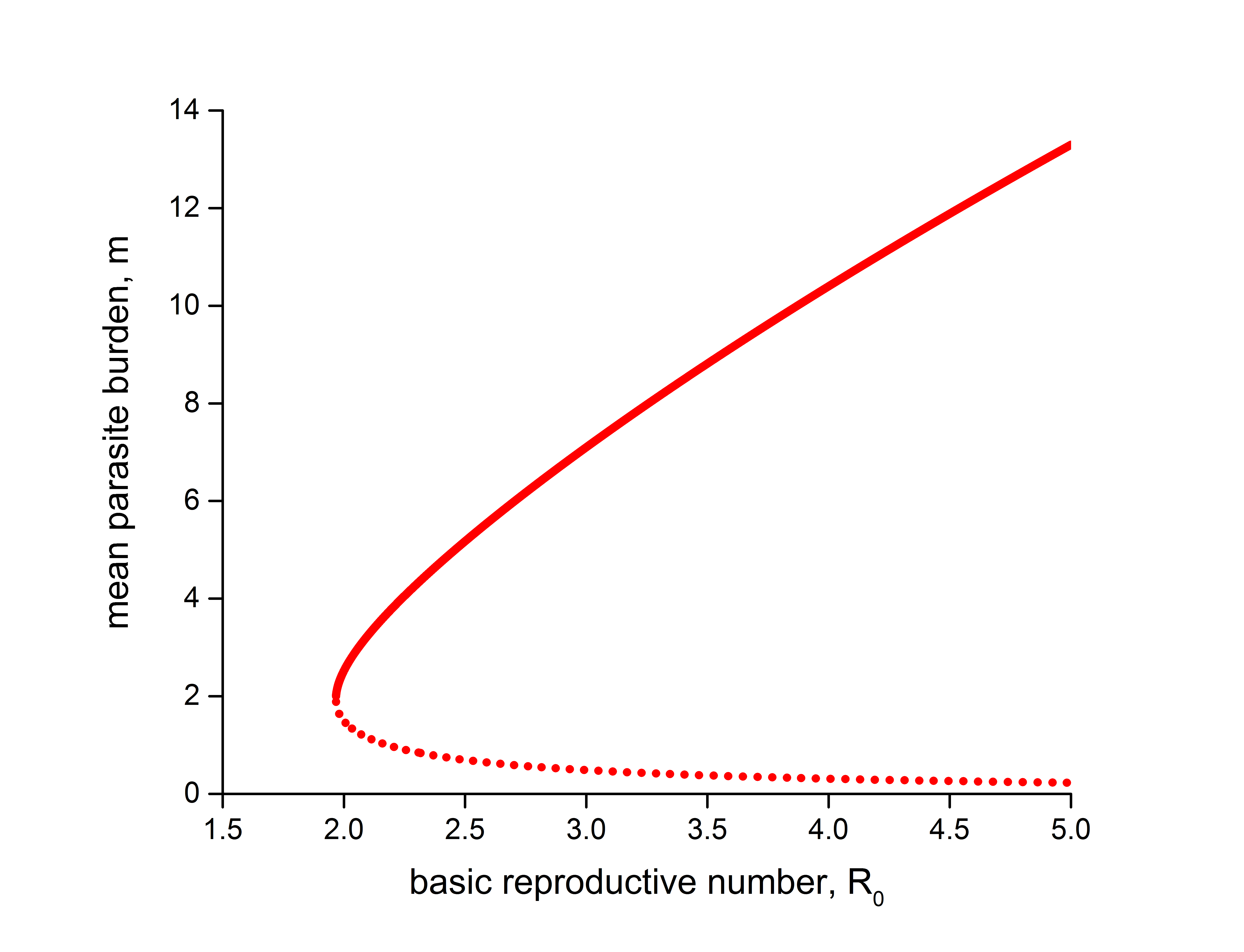}
%		\caption{In the graph we observe the three equilibria. The stable equilibrium at $M^*=$, the unstable equilibrium at $M^*=$ and the extinction equilibrium at $M^*=0$. Parameter values  }
		\caption{Bifurcation diagram generated by eq. \eqref{eqequilibrio}, parameter values $\alpha=0.5$, $k=0.7$ and $z=0.93$.
		The solid line and dotted line correspond to the stable and unstable branch, respectively, of the saddle-node bifurcation.}
		\label{f:phase}
	\end{figure}

	%{\color{green}
	%Realizando un análisis de bifurcación obtenemos que el
	%Debido  a este comportamiento el 
	%sistema formado por la ecuaciones (\ref{model1eq1}) y (\ref{model1eq2}) presenta una bifurcación nodo ensilladura en el punto $(M^*,R_0^*)$ donde estos valores están dados por
	By bifurcation analysis we obtain that the system of the  equations 
	\eqref{eqreservorio}-\eqref{eqm}
	%(\ref{model1eq1}) and (\ref{model1eq2}) 
	present a saddle-node bifurcation.
	The bifurcation point is $(\tilde m,\tilde R_0)$ where 
	\begin{equation}\label{meq}
	\begin{split}
	\tilde m=&\dfrac{k\left( \frac{1-\alpha z}{1-z}\right)^{\frac{1}{k+2}} - k}{(z-1)\left( \frac{1-\alpha z}{1-z}\right)^{\frac{1}{k+2}} + (1-\alpha z)}\\
	\tilde R_0=&\left[ \psi(\tilde m;k,z)\phi(\tilde m;k,z)\right]^{-1}
	\end{split}	
	\end{equation}
As shown in the next section, the system undergoes a saddle-node bifurcation and therefore, for  $R_0> \tilde R_0 $ there are three equilibria
 (see Figure \ref{f:phase}). 
	One of the solutions is the \textbf{stable}  endemic equilibrium %$\tilde m$ (\ref{meq}),  
	which is an attractor for a range of values of $R_0> \tilde R_0 $ .
	The other solution is an \textbf{unstable} equilibrium and corresponds to a repulsor in the phase plane, that is, a barrier where values of $m(t)$ above the unstable equilibrium are attracted towards the stable equilibrium and values of $m (t)$ below the unstable equilibrium are attracted to the \textbf{extinction} equilibrium $m^*= 0$, which is the trivial solution of the equation (\ref{eqMR0}) .
%	Due to the type of bifurcation present in the system, the solution of the equation \eqref{eqequilibrio} presents two solutions. 
%	The first is called \textbf{stable} and is the endemic solution of the system.
%	This equilibrium is an attractor for a range of values of $R_0> R_0 ^\star$ .
%	The second is known as \textbf{unstable} since it corresponds to a repulsor in the phase plane, that is, a barrier where the values of $M(t)$ above it are attracted to the stable equilibrium and the values of $M(t)$ below are attracted to the extinction equilibrium $M^*= 0$, which is the trivial solution of the equation (\ref{eqMR0}).
	%{\color{red} PONER REF LA FIGURA
	%}
	
	%{\color{red}
		To develop better control measures for macroparasitic diseases, it is necessary to know the relative importance of the different factors responsible for transmission.
		
		The transmission of macroparasitic diseases is related to the value of $R_0$. To predict which parameters have a higher impact on $R_0$, we must perform a sensitivity analysis on $R_0$.	
		
		The elasticity index o normalized sensitivity index measures the relative change of $R_0$ with respect to a parameter $x$, denoted by $\Gamma^{R_0}_{x}$ , and defined as
		\begin{equation}
		\Gamma^{R_0}_{x}=\dfrac{\partial R_0}{\partial x} \dfrac{x}{R_0}
		\end{equation}
		The sign of $\Gamma^{R_0}_{x}$ tells whether $R_0$ correlates positively or negatively  with the  parameter $x$; whereas its magnitude determines the relative importance of the parameter.
		
		For this model, the calculation of the elasticity indices are given by	
		%	The relative importance of each parameter for control measures can be estimated by computing elasticity indices. For this model,
		\begin{equation}
		\Gamma^{R_0}_{\lambda_0}=\Gamma^{R_0}_{\alpha}=\Gamma^{R_0}_{\rho}=1, 
		\qquad \Gamma^{R_0}_{\mu_h}=-\dfrac{\mu_h}{\mu_{h}+\mu_p},
		\qquad \Gamma^{R_0}_{\mu_p}=-\dfrac{\mu_p}{\mu_{h}+\mu_p}
		\end{equation}
		if $\frac{1}{\mu_h} \gg \frac{1}{\mu_p}$, then $\Gamma^{R_0}_{\mu_p}\approx -1$
		and 
		$\Gamma^{R_0}_{\mu_h}\approx 0$.
		
		Therefore the more sensitive parameters for $R_0$ are $\lambda_0$, $\alpha$, $\rho$ and $\mu_p$.
		However, $\lambda_0$ and $\alpha$ correspond to parameters related to the life-cycle of the parasite which are quite difficult to modify, so a control measure for macroparasitic diseases should target to the reduction of $\rho$ and/or the increase of $\mu_p$.
		
		Therefore, we can conclude from this analysis that the reduction of $R_0$ 
		is possible by reducing the egg contribution from the hosts to the reservoir, for example, by building latrines in the host community or by increasing parasite mortality, for example, through the application of periodic and specific antiparasitic treatments.

\subsection{Saddle-node bifurcation}\label{bifurcacion}
We will show that the basic model developed in the section \ref{ss:structure} presents a saddle-node bifurcation. Assuming the parasite reservoir at equilibrium (\ref{eqL})  the system reduces to the one-dimensional system 

\begin{equation*}
\dfrac{dm}{dt}=(\mu_h + \mu_p)\left[ R_0  \psi(m)\phi(m) -1 \right] m%\notag es para no tener numero en ecuacion
\end{equation*}
which we compactly denote by
$\dfrac{dm}{dt}=f(m,R_0)$.
A necessary condition for the existence of a saddle-node bifurcation at 
$(\tilde m,\tilde R_0)$ is

\begin{equation}
\begin{split}
f(\tilde m,\tilde R_0)&=0\qquad\\
\dfrac{\partial f}{\partial m}(\tilde m,\tilde R_0)&=0
\end{split}
\end{equation}
where the first of these conditions is the equilibrium condition \eqref{eqequilibrio} of the system
%donde la primera de estas condiciones es la condición de equilibrio (\ref{eqequilibrio}) del sistema 
\begin{equation*}
\psi(\tilde m;k,z)\phi(\tilde m;k,z)=1/\tilde R_0,
\end{equation*}
and so we get the following equilibrium condition for $\tilde m$
%y así obtenemos la siguiente condición para el equilibrio $m^{\star}$
\begin{equation}
\frac{\partial }{\partial m}\psi(\tilde m;k,z)\phi(\tilde m;k,z)=0	
\end{equation}
The value of $m$ corresponding to this last condition is
%El valor de $m$ correspondiente con esta ultima condición es 
\begin{equation}
\tilde m=\dfrac{k\left( \frac{1-\alpha z}{1-z}\right)^{\frac{1}{k+2}} - k}{-(1-z)\left( \frac{1-\alpha z}{1-z}\right)^{\frac{1}{k+2}} + (1-\alpha z)}	
\end{equation}
and its corresponding basic reproductive number is
%y el número reproductivo básico correspondiente es  
\begin{equation}
\tilde R_0=\left[ \psi(\tilde m;z,k)\phi(\tilde m;z,k)\right]^{-1}
\end{equation}	
 
A sufficient condition for the existence of a saddle-node bifurcation at $(\tilde m,\tilde R_0)$ is
\begin{equation}
\begin{split}
\dfrac{\partial f }{\partial R_0}(\tilde m,\tilde R_0)\neq0\\
\dfrac{\partial^2 f }{\partial m^2}(\tilde m,\tilde R_0)\neq0
\end{split}
\end{equation}

By a Taylor series expansion of the function $f$ in a neighborhood of $(\tilde m,\tilde R_0)$, the equation (\ref{eqMR0}) is left
\begin{equation}
{\scriptstyle	
	\frac{dm}{dt}=f(\tilde m,\tilde R_0)+(m-\tilde m)\frac{\partial f }{\partial m}\big\vert_{(\tilde m,\tilde R_0)}%\delta M
	+(R_0-\tilde R_0){\frac{\partial f }{\partial R_0}\big\vert_{(\tilde m,\tilde R_0)}}%\delta R_0
	+{\frac {1}{2}}(m-\tilde m)^2{\frac{\partial^2 f }{\partial m^2}}\big\vert_{(\tilde m,\tilde R_0)}%\delta M^{2}
	+\cdots 
}
\end{equation}
%Por lo tanto localmente en el punto $(m^{\star},R_0^{\star})$ la ecuación es de la forma
Therefore locally at the point $(\tilde m,\tilde R_0)$ the equation is of the form
\begin{equation}
\dfrac{dm}{dt}=A(R_0-\tilde R_0)+B(m-\tilde m)^2
\end{equation}
where the values $A=(\mu_h +\mu_p)\frac{\tilde m}{\tilde R_0}$ and $B=(\mu_h + \mu_p) R_0 \tilde m \frac{\partial^2 F}{\partial m^2}(\tilde m)$ with $F(m)= \psi(m,z,k)\phi(m,z, k)$
which is the normal form of a saddle-node bifurcation.

\section{A heterogeneous model}
In this section we will consider the more general and realistic case for a host population $H$. Unlike the homogeneous model presented in the previous section, here we present a model that accounts for host population heterogeneity, where subpopulations $H_i$ (e.g., age groups, risk groups, \cite{anderson1992infectious,anderson2014coverage,truscott2014modeling}) have different infection risks. The dynamics of infection for the case of a heterogeneous population is described as follows
\begin{equation}\label{model2}
	\begin{split}
		\dfrac{dm_i}{dt}&=\beta_i \ell - (\mu_h+\mu_p) m_i\\
		\dfrac{d\ell}{dt}&= 
		%\frac{\lambda_0}{2}    
		\lambda_0 \alpha
		\sum_i  N_i \rho_i  m_i F(m_i)   - (\mu_{\ell}+\sum_i \beta_i N_i ) \ell 
	\end{split}
\end{equation} 
where $N_{i}$ is the number of host in $H_i$.

\subsection{Equilibria and basic reproduction number} 
From the system \eqref{model2} we obtain that in equilibrium
%Del sistema \eqref{model2} obtenemos que en el equilibrio
%Suponiendo una situación de equilibrio para el reservorio $L$, obtenemos que 
\begin{equation}
	\ell^*=\frac{  \lambda_0 \alpha }{(\mu_{\ell}+\sum_i N_i \beta_i  )}   \sum_i \rho_{i} N_{i} m_{i} F(m_{i}) 
\end{equation} 
and substituting this in the rest of the equations of the initial system we obtain the following equation for the dynamics of the mean burden $m_{i}$ of the subpopulation $H_{i}$ of hosts
%y reemplazado esto en el resto de las ecuaciones del sistema inicial obtenemos las siguiente ecuación para la dinámica de la carga media $M_{ij}$ de la subpoblación $H_{ij}$ de hospedadores
\begin{equation}
	\begin{split}
		\dfrac{dm_{i}}{dt}=\beta_{i} \frac{\lambda_0\alpha}{ (\mu_{\ell}+\sum_j N_j \beta_j  ) }  
		\sum_j    N_i \rho_{j}  m_{j} F(m_{j})  - (\mu_h+\mu_p) m_{i}%\notag es para no tener numero en ecuacion
	\end{split}
\end{equation}
%La carga media del grupo $H_i=\sum_j H_{ij}$ de hospedadores %(con/sin servicios de $H_i$) 
%que denotaremos por $M_i$ viene dada por 
%\begin{equation}
%	M_i=\sum_j p_{ij} M_{ij} 
%\end{equation}
%%donde $p_{ij}$ es la proporción de $H_{ij}$ respecto de $H_i$ tal que $\sum_j p_{ij}=1$, 
%y su dinámica está descripta por 
%\begin{equation}
%	\begin{split}
	%		\dfrac{dM_i}{dt}=\left( \sum_j p_{ij}\beta_{ij}\right)  \frac{ \sigma \alpha \lambda}{\mu_L}  \sum_i \pi_i  \sum_j \rho_{ij} p_{ij} M_{ij} F(M_{ij})  - (\mu_H+\mu_W) M_i%\notag es para no tener numero en ecuacion
	%	\end{split}
%\end{equation}
%La carga media $m$ de la población total de hospedadores $H=\sum_i H_i$ esta dada por
The mean burden $m$ of the total host population $H=\bigcup_i H_i$ is given by
\begin{equation}
	m=\sum_i \pi_i m_{i} 
\end{equation}
where $\pi_i$ is the portion of the population $H$ corresponding to the subpopulation $H_i$, and which is described by
%donde $\pi_i$ es la porción de la población $H$ correspondiente a la subpoblación $H_i$, y la cual está descripta por 
\begin{equation}
	\begin{split}
		\dfrac{dm}{dt}= \left( \sum_i N_i \beta_{i} \right)  
		\frac{ \lambda_0 \alpha }{(\mu_{\ell}+\sum_j N_j \beta_j  )}  
		\sum_j \rho_{j} \pi_{j} m_{j} F(m_{j})   -(\mu_{h}+\mu_p) m%\notag es para no tener numero en ecuacion
	\end{split}
\end{equation}
%Suponiendo el diferencial en cero
%de esta ecuación, la carga media de parásitos en equilibrio, $m^*$, para la población total viene dada por
From this equation, the equilibrium mean parasite burden, $m^*$, for the total population is given by
\begin{equation}
	\sum_i \pi_i \frac{ \lambda_0 \alpha \rho_{i}}{ (\mu_{\ell}+\sum_j N_j \beta_j  )(\mu_{h}+\mu_p)} 
	\left( \sum_j N_{j} \beta_{j} \right) F( m^*_{i}) m^*_{i} - m^*=0 
\end{equation}
%Esta no es una expresión explícita de los equilibrios $m_{i}^*$. Por lo tanto, el valor de los equilibrios solo se pueden resolver numéricamente. 
%Una condición de equilibrio para las cargas medias de cada subpoblación $H_{i}$ viene dada por 
This is not an explicit expression of the equilibria $m_{i}^*$. Therefore, the equilibrium value can only be solved numerically.
An equilibrium condition for the mean burdens of each subpopulation $H_{i}$ is given by
\begin{equation}%\label{eqequilibrio}
	F(m^*_{i})=1/R_0^{i}
\end{equation}
%	\begin{equation}
	%	R_{0\rho}=\frac{\sigma \alpha \lambda }{ \mu_L (\mu_H+\mu_M)} (\beta \pi + \hat\beta \hat\pi)\rho
	%	\end{equation}
%	donde $R_{0\hat\rho}$ y $R_{0\rho}$ es la contribución del grupo provisto y no provisto de agua y saneamiento respectivamente, para más detalles ver Apéndice (poner referencia).
%	
%	donde $\beta_1=\beta$, $\beta_2=\beta(1-\epsilon_A)$, $\rho_1=\rho$ y $\rho_2=\rho(1-\epsilon_S)$.
%donde definimos por 
%$R_0^{i}=\frac{ \lambda_0 \rho_{i}}{2 (\mu_{\ell}+\sum_j \pi_j \beta_j  )(\mu_{h}+\mu_p)} \left( \sum_j \pi_j\beta_{j} \right) $ 
%al número reproductivo básico propio de cada subpoblación $H_{i}$ que es el número de hembras adultas que surgen 
%%en la población total del hospedador 
%de una hembra adulta de un hospedador de la subpoblación $H_{i}$ en ausencia de los efectos de la denso-dependencia y la probabilidad de apareamiento. 
%Para esta situación de equilibrio obtenemos que la carga media de parásitos de cada subpoblación $H_{i}$ viene dada por $m_{i}^*=\frac{\beta_{i}}{ \sum_j \pi_j\beta_{j} }m^*$.
%El número reproductivo básico general $R_0$ para la población total esta dado por %\cite{diekmann2012mathematical}
where we define by
$R_0^{i}=\frac{ \lambda_0 \alpha \rho_{i}}{ (\mu_{\ell}+\sum_j N_j \beta_j )(\mu_{h}+\mu_p)} \left( \sum_j N_j\beta_{j} \right) $
to the basic reproductive number of each subpopulation $H_{i}$
which is the number of adult females 
that are born of a 
adult female from a host in subpopulation $H_{i}$ in the absence the effects of density-dependence and the mating probability.
Note  what for a large $N$ value $R_0^i\approx \frac{ \lambda_0 \alpha  \rho_i }{ (\mu_h + \mu_p) }$.  
Also for this equilibrium situation, we obtain that the mean parasite burden of each subpopulation $H_{i}$ is given by $m_{i}^*=\frac{\beta_{i}}{ \sum_j \pi_j\beta_{j } }m^*$.

The general basic reproductive number $R_0$ for the total population is given by %\cite{diekmann2012mathematical}
\begin{equation}\label{valorR0}
	R_{0}=\frac{\lambda_0 \alpha }
	{ (\mu_{\ell}+\sum_j N_j \beta_j  )(\mu_{h}+\mu_p)}
	\sum_i N_i \rho_{i} \beta_{i}   
\end{equation}
%donde suponemos la ausencia de los efectos de la denso-dependencia y la probabilidad de apareamiento \citep{anderson1992infectious}, es decir suponemos en el sistema (\ref{eqmodel2})  la función $F$ igual a la unidad.
where we assume the absence the effects of density-dependence and the mating probability \cite{anderson1992infectious}, that is, we assume in the system \eqref{model2} the function $F$ equal to unity.
A relationship between $R_0$ and $R_0^i$ is given by
\begin{equation}
R_{0}=\frac{\sum_i N_i\beta_{i}R_0^i}
{\sum_j N_j \beta_{j}}   
\end{equation}
therefore we get that $\min R_0^i\leq R_0 \leq \max R_0^i$
, then we can interpret to $R_0$ as an average value of the $R_0^i$.

\section{Discussion and Conclusions}
%{\color{red}
In this work, we developed deterministic mathematical models for the transmission dynamics of macroparasite infections. 

We show how fundamental parameters related to production of fertilized parasites eggs are estimated from statistical models for the distribution of
parasites within hosts.	

We considered both homogeneous and heterogeneous host communities. 
The analyzed models show that the basic reproduction number $R_0$ strongly depends 
on the %{\color{red}eggs?} 
host egg contributions to the reservoir (which depend of the parameters $\rho$, $\alpha$, and the parasite fecundity at low densities $\lambda_0 $), and on the parasite mortality ($\mu_p$). 
Therefore, to achieve a reduction in $R_0$ we must, for example, build latrines in the host community or implement regular and specific antiparasitic treatments.

%Therefore, to achieve a reduction in $R_0$, we must 
%%reduce host egg contributions to the reservoir, 
%for example by building latrines in the host community or 
%%Another reduction of $R_0$ is through the increase in parasite mortality, for example 
%by the implementation
%of regular and targeted antiparasitic treatments.

For the homogeneous model we present a bifurcation analysis and show that this model exhibits a saddle-node bifurcation.
The bifurcation parameter depends on the functions $\psi$ and $\phi$ which in turn depend on the assumed
distribution of parasites (see \cite{lopez2022general}).

More refined models may be developed from the simple models presented here which may be useful in the design and evaluation of different control strategies.

\section*{Aknowledgements}
This work was partially supported by grant CIUNSA 2018-2467. JPA is a member of the CONICET. GML is a doctoral fellow of CONICET.

\section*{Conflict of Interest}
The authors have declared no conflict of interest.

%Reference

\bibliographystyle{apa}
\bibliography{biblio}	
\end{document}